\documentclass[aps,showpacs,showkeys,preprint]{revtex4}
\usepackage[dvips]{graphicx}
\usepackage{latexsym}

\begin{document}

\preprint{\vbox{\hbox{\tt gr-qc/0608032}}}

\title{Accelerating Universe in two-dimensional noncommutative dilaton cosmology}
\author{Wontae Kim}
  \email{wtkim@sogang.ac.kr}
  \affiliation{Department of Physics, Center for Quantum Spacetime,
    and Basic Science Research Institute,
    Sogang University, C.P.O. Box 1142, Seoul 100-611, Korea}
\author{Myung Seok Yoon}
  \email{yoonms@sejong.ac.kr}
  \affiliation{Department of Physics, Sejong University, Seoul 143-747, Korea}

\date{\today}

\begin{abstract}
   We show that the phase transition from the decelerating universe
   to the accelerating universe, which is of relevance to the
   cosmological coincidence problem, is possible in the
   semiclassically quantized two-dimensional dilaton gravity by
   taking into account the noncommutative field variables during the
   finite time. Initially, the quantum-mechanically induced energy
   from the noncommutativity among the fields makes the early
   universe decelerate and subsequently the universe is
   accelerating because the dilaton driven cosmology becomes dominant
   later.
\end{abstract}


\maketitle

\section{Introduction}
\label{sec:intro}

It has been proposed that the discovery of the accelerating universe
from the observations of the supernovae~\cite{perlmutter}, is
intriguingly related to the dark energy~\cite{spergel}. There may be
many candidates for the dark energy described by the equation of state
parameter, which is defined as the ratio of pressure to energy
density, $w (\equiv p/\rho) < -1/3$, responsible for the accelerating 
universe. If it is even more exotic, like the phantom field of
$w < -1$~\cite{caldwell}  in order for compensating the ordinary matter,
the simplest realization is to take the wrong-sign kinetic term
violating the dominant energy condition. The quantum gravity effect
for the phantom, scalar tensor theory, and the other interesting
models have been well appreciated in Refs.~\cite{carroll,bbmm,no,klm,wc}. The
ordinary matter in the Friedman equation based on the Einstein theory
gives rise to the decelerating phase of the universe while the dilaton
gravity from the low energy string theory presents the expected
accelerating universe since the dilaton plays the role of the phantom
field. However, the two representative models just maintain
their own phases once they are determined by the matter contents.

On the other hand, the two-dimensional dilaton gravity is very useful
in studying the classical and quantum aspects~\cite{mrcm} because it
has fewer degrees of freedom and is free from the renormalizability
problem rather than the four-dimensional counterpart. So, in this
simple context, the phase transition from the accelerating universe to
the decelerating FRW phase called the graceful exit problem has been
extensively studied in terms of the quantum back reaction of the
geometry in Refs.~\cite{gv,rey,bk,ky:dg,ky:bd}. In these models,
the curvature scalar proportional to the acceleration of the scale
factor has a definite sign which never changes its sign in these
models. Recently, the transition is demonstrated by the numerical
method in the two-dimensional cosmology by introducing the van der
Waals equation of state instead of the usual perfect cosmic
fluid~\cite{cdkz}.
 
In this paper, we would like to present an exactly soluble model
showing the phase change from the decelerating universe to
accelerating universe by using the well-known two-dimensional dilaton
gravity~\cite{cghs,rst,bpp,kv} without assuming any classical matter
contents. So, if it can happen, the phase change may come from the
nontrivial time-dependent vacuum state. However, in the ordinary
two-dimensional dilaton cosmology, the nontrivial vacuum does not
appear even in the quantized theory. Therefore, for this purpose, we shall
assume the nontrivial Poisson brackets between fields similar to the
noncommutative algebra in Ref.~\cite{sw}. As a matter of fact, the
deformed brackets in the homogeneous spacetime generate new
equations of motion involving the noncommutative
parameter~\cite{bn,vas,ko}, which will be defined within the finite
time. Initially, the quantum-mechanically induced positive energy from
the noncommutativity between the fields makes the universe
decelerate and subsequently the universe is accelerating because the
dilaton cosmology becomes dominant eventually, where the dilaton field
as a dark energy source causes an acceleration~\cite{bbmm}. 

We will recast the commutative variant of a dilaton
model. In this model, it will be shown that the only accelerating universe
is possible irrespective of any vacuum states. 
Then, in the noncommutative dilaton cosmology, the
modified Poisson algebra gives the new set of equations of motion and
constraint equations, which yields the nontrivial vacuum energy
density depending on the noncommutative parameter and gives desired the
phase change of the universe. Finally, we discuss and summarize our results.


We now start with the following dilaton gravity action, 
\begin{equation}
   S = S_{\rm D} + S_{\rm cl} + S_{\rm qt}, \label{act:total}
\end{equation}
where the classical dilaton action from the low-energy string theory is 
\begin{equation}
  S_{\rm D} = \frac{1}{2\pi} \int d^2 x \sqrt{-g} e^{-2\phi} \left[
     R + 4 (\nabla \phi)^2 + 4 \lambda^2\right], \label{act:dg}
\end{equation}
and the classical matter and its quantum correction are given as
\begin{eqnarray}
  S_{\rm cl} &=& - \frac{1}{2\pi} \int d^2 x \sqrt{-g} \frac12
     \sum_{i=1}^{N} (\nabla f_i)^2, \label{action:cl} \\
  S_{\rm qt} &=& \frac{\kappa}{2\pi} \int \sqrt{-g} \left[ 
     - \frac14 R\frac{1}{\Box} R + (\nabla\phi)^2 - \phi R \right],
     \label{action:qt}
\end{eqnarray}
where $\kappa=(N-24)/12$ and the cosmological constant $\lambda^2$
sets to be zero. The first term in Eq.~(\ref{action:qt}) comes from
the Polyakov effective action of the classical matter
fields~\cite{cghs,rst} and the other two local terms are introduced
in order to solve the semi-classical equations of motion
exactly~\cite{bpp}. The higher order of quantum correction beyond the
one-loop is negligible in the large $N$ approximation where $N \to
\infty$ and $\hbar \to 0$, so that $\kappa$ is assumed to be positive
finite constant. 

In the conformal gauge, $ds^2 = -e^{2\rho} dx^+ dx^-$, the total
action and the constraint equations are written as
\begin{eqnarray}
 S &=& \frac{1}{\pi} \int\/ d^2 x \bigg[ e^{-2\phi}\left(
       2\partial_+\partial_-\rho - 4\partial_+\phi\partial_-\phi
       \right) - \kappa\big( \partial_+\rho\partial_-\rho
       + 2\phi\partial_+\partial_-\rho \nonumber \\
   & & + \partial_+\phi\partial_-\phi \big) + \frac12 \sum_{i=1}^{N}
       \partial_+f_i\partial_-f_i \bigg] \label{act:conf}
\end{eqnarray}
and
\begin{eqnarray}
  & & e^{-2\phi}\left[ 4\partial_\pm\rho\partial_\pm\phi
      - 2\partial_\pm^2\phi \right]
      + \frac12\sum_{i=1}^N\left(\partial_\pm f\right)^2
      + \kappa\left[ \partial_\pm^2\rho
      - \left(\partial_\pm\rho\right)^2\right] \nonumber \\
  & & \qquad\qquad\qquad - \kappa\left(
      \partial_\pm^2\phi - 2\partial_\pm\rho\partial_\pm\phi\right)
      - \kappa\left(\partial_\pm\phi\right)^2
      - \kappa t_\pm = 0, \label{constr:conf}
\end{eqnarray}
where $t_{\pm}$ reflects the nonlocality of the induced gravity of the
conformal anomaly. Note that our semiclassical action (5) is defined 
by the one-loop quantum correction of the classical matter action (3) 
which is described by the Polyakov nonlocal action along with the two 
local ambiguity terms in Eq. (4). In fact, the dilaton-gravity part
(1) is not quantized so that the total action is partially quantized, 
which means that we will treat the so-called semiclassical action. 
Then, we can study the back reaction of the geometry due to the quantized matter.

Without the classical matter, $f_i = 0$, defining
new fields as $\Omega = e^{-2\phi}$, $\chi = \kappa (\rho
-\phi) + e^{-2\phi}$~\cite{bpp,bc}, the gauge fixed action is obtained
in the simplest form of 
\begin{equation}\label{act:new}
S = \frac{1}{\pi} \int\/d^2 x \left[
    \frac{1}{\kappa}\partial_+\Omega\partial_-\Omega 
    - \frac{1}{\kappa}\partial_+\chi\partial_-\chi \right]
\end{equation}
and the constraints are given by
\begin{equation}
\kappa t_\pm = \frac{1}{\kappa}\left( \partial_\pm\Omega\right)^2 
    - \frac{1}{\kappa} \left(\partial_\pm\chi\right)^2 
    + \partial_\pm^2\chi. \label{constr:new}
\end{equation}

In the homogeneous spacetime, the Lagrangian and the constraints are
obtained as
\begin{eqnarray}
  L &=& \frac{1}{4\kappa} \dot{\Omega}^2 
        - \frac{1}{4\kappa} \dot{\chi}^2, \label{L} \\
  \frac{1}{4\kappa} \dot{\Omega}^2 \!&-&\! \frac{1}{4\kappa}
        \dot{\chi}^2 + \frac14 \ddot{\chi} - \kappa t_{\pm}=0, 
        \label{con}
\end{eqnarray}
where the action is redefined by $S/L_0 = \frac{1}{\pi} \int dt L$ and
$L_0=\int dx$, and the overdot denotes the derivative with respect to
the cosmic time $t$. Then, the Hamiltonian becomes
\begin{equation}
  \label{H}
  H = \kappa P_\Omega^2 - \kappa P_\chi^2
\end{equation}
in terms of the canonical momenta $P_\chi = - \frac{1}{2\kappa}
\dot{\chi}$, $P_\Omega = \frac{1}{2\kappa} \dot{\Omega}$.

Let us now define the nonvanishing Poisson brackets,
\begin{equation}
  \label{PB:com}
  \{\Omega, P_\Omega\}_{\mathrm{PB}} = \{\chi, P_\chi\}_{\mathrm{PB}} =1,
  \quad \mathrm{others} = 0
\end{equation}
and the Hamiltonian equations of motion in~Ref.~\cite{bk:H} are given by
$\dot {\cal O}  = \{ {\cal O}, H \}_{\mathrm{PB}}$ where ${\cal O}$
represents fields and corresponding momenta, then they are explicitly
written as
\begin{eqnarray}
   & & \dot\chi = - 2\kappa  P_\chi, \quad \dot\Omega = 2\kappa
       P_\Omega, \label{1st:x_com} \\
   & & \dot{P}_\chi =0, \quad \dot{P}_\Omega = 0. \label{1st:p_com}
\end{eqnarray}
Since the momenta $P_\Omega$ and $P_\chi$ are constants of motion as
seen from Eq.~(\ref{1st:p_com}), we easily obtain the solutions as
\begin{eqnarray}
  & & \Omega =  2\kappa P_{\Omega_0} t + A_0, 
      \label{sol:Omega_com} \\
  & & \chi =  -2\kappa P_{\chi_0} t + B_0,
      \label{sol:chi_com}
\end{eqnarray}
where $ P_\Omega = P_{\Omega_0}$, $P_\chi = P_{\chi_0}$, $A_0$, and
$B_0$ are arbitrary constants. Next, the dynamical
solutions~(\ref{sol:Omega_com}) and (\ref{sol:chi_com}) should satisfy
the constraint~(\ref{con}),
\begin{equation}
  \label{constr:com}
  \kappa t_\pm = \kappa (P_{\Omega_0}^2 - P_{\chi_0}^2),  
\end{equation}
which is related to the vacuum energy density~\cite{cghs,rst,bpp}
\begin{eqnarray}
  <T_{\pm\pm}> &=& -\kappa t_\pm - \kappa (P_{\Omega_0} - P_{\chi_0})^2
              \nonumber \\
      &=& -2\kappa P_{\Omega_0} ( P_{\Omega_0} + P_{\chi_0}).
  \label{energy:com}
\end{eqnarray}
In this model, the quantum-mechanically induced vacuum energy is, at
the best, constant. 
Essentially, the Hamiltonian and the boundary functions $t_\pm$ are
different 
in that the latter is just a part of constraint equations. The
solutions from the 
equations of motion should satisfy the constraint equations. 
The boundary functions can be in general time dependent depending on 
the choice of the matter states semiclassically whereas the
Hamiltonian 
is time independent. In this model, they happen to be the same form, 
however as seen from Eq. (17), $t_{\pm}$ is composed of 
the integration constants instead of the dynamical variables in Eq. (11).
 
On the other hand, by using Eqs.~(\ref{sol:Omega_com}) and
(\ref{sol:chi_com}),
the curvature scalar is calculated as
\begin{equation}
  \label{R:com}
  R = 4\kappa^2 P_{\Omega_0}^2 e^{-2\rho + 4\phi} 
    = 4 \kappa^2 P_{\Omega_0}^2 \frac{e^{ -2B_0 +4 \kappa P_{\chi_0} 
      t} }{ A_0 + 2\kappa P_{\Omega_0} t }. 
\end{equation}
Since $\Omega=e^{-2\phi}$ in Eq.~(\ref{sol:Omega_com}) should be
positive definite, there are two types of branches: the first one is
that $t> -A_0/(2\kappa P_{\Omega0})$ for the positive charge of
$P_{\Omega_0}>0$, and the second is that $t<A_0/(2\kappa
P_{\Omega0})$ for the negative charge of $P_{\Omega_0}<0$. Note that
the universe is always accelerating irrespective of the vacuum energy
density since the expression for the curvature
scalar in Eq.~(\ref{R:com}) is written as $R=2\ddot{a}/a$ in the comoving coordinates, $ds^2
= -d\tau^2 + a^2(\tau) dx^2$, where $a(\tau)$ is a scale factor. 

The curvature singularity corresponding to the infinite acceleration
appears at $t \rightarrow - A_0/(2\kappa
P_{\Omega0})$ while there exists another
singularity at $t \rightarrow \infty$ for $P_{\chi_0} > 0$.
In the next section, we shall choose the former case of $P_{\chi_0}
<0$ to avoid the infinite acceleration in the future and to obtain
the regular geometry, although it is singular at the one 
instant $t \rightarrow - A_0/(2\kappa P_{\Omega0})$. However, this
singularity becomes unimportant since this geometry will not be used beyond
the singularity. 
  
The standard lore tells us that the ordinary matter causes the
decelerating universe, however, in our case, the effect of the dilaton
which has wrong sign kinetic term survives the induced energy
Eq.~(\ref{energy:com}) and it seems to be much more dominant whatever
the signature of induced energy is. Thus, in this accelerating model, we
are tempted to have a quantum-mechanically induced positive energy in
the early universe which may moderate the harsh acceleration.


Now, we study whether the phase change of the universe is
possible or not in the context of the noncommutative algebra. So, we
will consider the modified Poisson brackets corresponding to the
noncommutative algebra \cite{sw,bn},
\begin{eqnarray}
  & & \{ \Omega, P_\Omega \}_{\mathrm{MPB}} = 
      \{ \chi, P_\chi \}_{\mathrm{MPB}} = 1, \nonumber \\ 
  & & \{ \chi, \Omega \}_{\mathrm{MPB}} = 
      \theta_1 [\epsilon(t-t_1) - \epsilon(t-t_2)], \nonumber \\
  & & \{ P_\chi, P_\Omega \}_{\mathrm{MPB}} = 
      \theta_2 [\epsilon(t-t_1) - \epsilon(t-t_2)], \nonumber\\
  & & \mathrm{others} = 0, \label{PB:non}
\end{eqnarray}
where $\theta_1$ and $\theta_2$ are two independent positive
constants, and $\epsilon(t)$ is a step function, $1$ for $t>0$ and
$0$ for $t<0$. Thus, these are nontrivial and $\theta$-dependent for
the finite time interval of $t_1 < t < t_2$ compared to
the ordinary brackets. For $t> t_2$, they recover
Eq.~(\ref{PB:com}). 
The two parameters are independent of the Plank constant and 
we do not  intend to perform one more quantization of the
semiclassical 
action. These constants are just assumed parameters in order to obtain 
the desired result. Of course, depending on models, they can be
derived 
from the classical constraint analysis. For example, the nontrivial 
Poisson algebra between the momenta can be obtained from the
model of a very slowly moving charged particle in the constant
magnetic field. The conventional Poisson algebra are modified 
by the constraint which yields nontrivial Poisson algebra 
proportional to the constant magnetic field  in terms of the 
classical Hamiltonian constraint analysis [23]. However, 
in our model, we just assume the noncommutative parameters as an ansatz. 

Using the Hamiltonian~(\ref{H}), 
for $t_1 < t < t_2$,
the previous
equations of motion are promoted to the followings,
\begin{eqnarray}
  & & \dot\chi = \{ \chi, H \}_{\mathrm{MPB}} = -2\kappa P_\chi, \quad
      \dot\Omega =  \{ \Omega, H \}_{\mathrm{MPB}} = 2\kappa P_\Omega, 
      \label{1st:x_non}\\
  & & \dot{P}_\chi =  \{ P_\chi, H \}_{\mathrm{MPB}} =2\kappa \theta_2
      P_\Omega, \quad \dot{P}_\Omega = \{ P_\Omega, H \}_{\mathrm{MPB}} 
      = 2\kappa\theta_2 P_\chi. \label{1st:p_non}
\end{eqnarray}
This is a definitely effective modification at the semiclassical level 
for the finite time interval because the original semiclassical
equation of motion 
(13) and (14) are reproduced if the noncommutative parameters vanish. 
The first order equations of motion (13) and (14) in 
the Hamiltonian formulation are in fact the same with 
the Euler-Lagrangian equations of motion from the semiclasscial 
action. So, the modified equations of motion (21) and (22) 
are nothing but the simiclasscial equations of motion 
which are just improved by the modified Poisson brackets. 
Our assumption for the Poisson brackets (20) does not mean that they are quantum commutators. 

Note that the momenta are no more constants of motion because of
nonvanishing $\theta_2$, hereby, a new set of equations of motion from
Eqs.~(\ref{1st:x_non}) and (\ref{1st:p_non}) are obtained,
\begin{equation}
  \label{eom:non}
  \ddot\chi = -2\kappa \theta_2 \dot\Omega, \quad \ddot\Omega =
  -2\kappa \theta_2 \dot\chi.
\end{equation}
We have introduced $\theta_1$ without loss of generality. However, it plays
no role in our calculations 
because the Hamiltonian does not have any fields but it has only
momenta. To affect the equations of motion, 
the Hamiltonian should have field components since $\theta_1$ is a 
result of correlation among the fields.

The solutions for the above coupled equations of motion are easily
solved as
\begin{eqnarray}
  \Omega &=& \alpha e^{-2\kappa\theta_2 t} + \beta
            e^{2\kappa\theta_2 t} + A, \label{sol_gen:Omega_non} \\
  \chi &=& \alpha e^{-2\kappa\theta_2 t} - \beta
            e^{2\kappa\theta_2 t} + B, \label{sol_gen:chi_non}
\end{eqnarray}
where $\alpha$, $\beta$, $A$, and $B$ are constants, and they should
satisfy the constraint equation~(\ref{con}),
\begin{equation}
  \label{constr_gen:non}
  \kappa t_\pm = \kappa^2 \theta_2^2 (\alpha e^{-2\kappa\theta_2 t} 
      -\beta e^{2\kappa\theta_2 t}) -4\kappa\theta_2^2 \alpha\beta,
\end{equation}
which determines the unknown time-dependent function $t_\pm$. 

Now, taking $\beta=-\alpha>0$, the solutions and the boundary functions $t_\pm$
are written as
\begin{eqnarray}
  \Omega &=& e^{-2\phi} = 2 \beta \sinh (2\kappa\theta_2 t) + A,
     \label{sol:Omega_non} \\
  \chi &=& \kappa(\rho - \phi) + e^{-2\phi} =  - 2 \beta \cosh
     (2\kappa\theta_2 t) + B, \label{sol:chi_non}
\end{eqnarray}
and
\begin{equation}
  \label{constr:non}
  \kappa t_\pm = -2 \beta \kappa^2 \theta_2^2 \cosh (2\kappa\theta_2
  t) + 4\kappa\beta^2\theta_2^2,
\end{equation}
respectively. Then, the induced vacuum energy is obtained as
\begin{equation}
  \label{energy:non}
  <T_{\pm\pm}> = -\kappa t_\pm - 2\beta \kappa^2 \theta_2^2
    e^{2\kappa\theta_2 t} - 4\beta^2\kappa\theta_2^2
    e^{4\kappa\theta_2 t}.
\end{equation}
Note that among the two positive constants, the only
$\theta_2$ plays an important role in our analysis. Furthermore,
similarly to the previous commutative case, $\Omega = e^{-2\phi}$ in
Eq.~(\ref{sol:Omega_non}) is positive definite, so that the initial
time should be restricted to $t_1 > - [1/(2\kappa\theta_2)] \sinh^{-1}
[A/(2\beta)]$. Especially, for $A=0$, the time interval become $0<t_1
< t < t_2$. Hereafter, we regard $t_1$ as  the initial time of the
beginning of the universe in our model. 

\begin{figure}[pt]
  \includegraphics[width=0.6\textwidth]{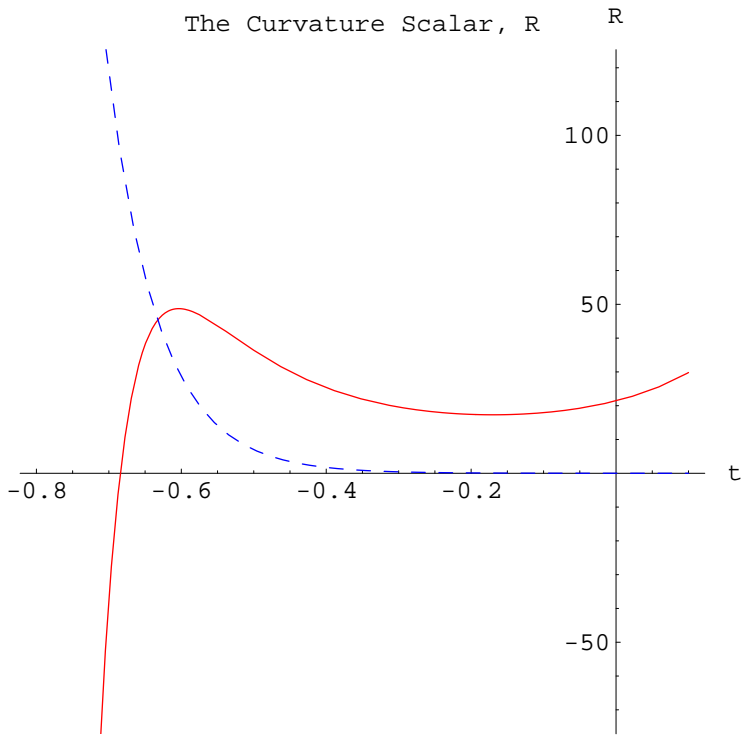}
  \caption{The solid line shows that the $\theta_2$-dependent
          curvature scalar in Eq.~(\ref{R:non}) is changing from the
          negative to the positive region while the dashed line for
          Eq.~(\ref{R:com}) for $P_{\chi_0}< 0$ is always accelerating
          and converges at $t \to \infty$. And, the solid line and
          dashed line intersects each other at $t=t_2$. This figure is
          plotted in the case of $\beta=\kappa=\theta_2=1$, $A=10$,
          and $B=3$, $t>t_1$. Then, $t_1 \approx -1.156$ and $t_2
          \approx -0.632 $ and the consistent constants satisfying the
          continuity equations are chosen as $P_{\Omega_0} \approx
          3.826$, $P_{\chi_0} \approx -3.262$, $A_0 \approx 11.579$,
          and $B_0 \approx 3.301$.} 
  \label{fig:R}
\end{figure}

At this juncture, from Eqs.~(\ref{sol:Omega_non}) and
(\ref{sol:chi_non}), we calculate the curvature scalar related to the
acceleration and deceleration in terms of $R=2\ddot{a}/a$ in the
comoving coordinates, then
\begin{eqnarray}
  R_{\theta} &=& -8 \beta \kappa^2\theta_2^2 \frac{\exp(4\beta %
        \cosh(2\kappa\theta_2 t) -2B)}{2\beta \sinh (2\kappa %
        \theta_2 t) + A} \bigg[ \cosh(2\kappa\theta_2 t) \left(
        2\beta e^{2\kappa\theta_2 t} + A \right) \nonumber \\
  & &  - \frac{2}{\kappa} e^{2\kappa\theta_2 t} \left(
        2\beta\sinh(2\kappa\theta_2 t) + A\right) \left(
        2\beta\sinh(2\kappa\theta_2 t) + A + \frac{\kappa}{2}\right)
        \bigg]. \label{R:non}
\end{eqnarray}
It is of interest to note that the sign of the curvature scalar is
remarkably changing from the negative to the positive region for
$t_1<t<t_2$ as seen from Fig.~\ref{fig:R}, which is reminiscent of the
evolution of the recently observed accelerating universe from the
decelerating universe. As shown in Fig.~\ref{fig:R}, the solid line
shows that the decelerating universe evolves into the accelerating
universe and eventually it turns out that it is infinitely
accelerating at $t_2 \to \infty$, however, it is unnatural to
consider this case. So, one might think that just after $t_2$ the
aforementioned regular accelerating geometry~(\ref{R:com}) can be
patched up this geometry. Note that the acceleration in the previous
pure accelerating geometry converges to zero for $t \to \infty$ for
 $P_{\chi_0}< 0$.

Intuitively, it is plausible to assume that the
extraordinary modified Poisson brackets are not allowed in the present
large universe, which implies that we should consider the normal Poisson
brackets yielding the original commutative geometry. Then, it is clear
that if we set the initial time $t=t_1$, then the final time $t_2$ to
suspend the noncommutativity should be located at the positive region
of the scalar curvature as far as the universe is connected with the
regular accelerating cosmology.

Therefore, let us now describe the geometry from the decelerating
universe to the accelerating universe which finally ends up with the
vanishing curvature scalar corresponding to the zero
acceleration. Then, our two different solutions should be patched at
$t=t_2$ to obtain the transition from noncommutative cosmology to
commutative one. So, matching the solutions~(\ref{sol:Omega_non}) and
(\ref{sol:chi_non}) with Eqs.~(\ref{sol:Omega_com}) and
(\ref{sol:chi_com}), and their time derivatives are also continuous at
$t=t_2$ yield the following conditions, 
\begin{eqnarray}
  \beta &=& \frac{P_{\Omega_0}}{2\theta_2} {\rm sech}\, (2\kappa\theta_2
            t_2), \label{patch:beta} \\
  A &=& A_0 + \frac{P_{\Omega_0}}{\theta_2} \left[ 2\kappa\theta_2 t_2 -
            \tanh (2\kappa\theta_2 t_2) \right], \label{patch:A} \\
  B &=& B_0 - \frac{P_{\chi_0}}{\theta_2} \left[ 2\kappa\theta_2 t_2 -
            \coth (2\kappa\theta_2 t_2) \right], \label{patch:B} \\
  \frac{P_{\chi_0}}{P_{\Omega_0}} &=& \tanh (2\kappa\theta_2
            t_2). \label{patch:chi} 
\end{eqnarray}
There are in fact 8-independent constants, however, from these
matching conditions and the relation of $\beta = -\alpha$, and the
time translational symmetry, the resulting independent number of
constants is $8-(4 +1 +1) =2$. For example, the independent variables
may be chosen as $P_{\Omega_0}$ and $A_0$, conveniently. 

Next, we
assign one more condition of $R(t_2)=R_{\theta}(t_2)$ in order to find
out the appropriate time ``$t_2$'' which connects the respective
scalar curvatures. This continuity requirement leads to
\begin{eqnarray}
 P_{\Omega_0}^2 &=& 2\beta\theta_2^2 \bigg[ \frac{2}{\kappa}
    e^{2\kappa\theta_2 t_2} \left( 2\beta\sinh (2\kappa\theta_2 t_2) 
    + A \right) \left( 2\beta\sinh (2\kappa\theta_2 t_2) + A 
    + \frac{\kappa}{2} \right) \nonumber \\
 & & - \cosh (2\kappa\theta_2 t_2) \left( 2\beta e^{2\kappa\theta_2
    t_2} + A \right)\bigg], \label{patch:R}
\end{eqnarray}
which corresponds to requirement that up to the second derivatives of
the metric and dilaton fields are continuous. From the beginning, we
consider that $\kappa$, $\theta_2$, and $P_{\Omega_0}$ are positive
and $P_{\chi_0}$ is negative, then from Eq.~(\ref{patch:chi}) the
consistent patching appears at the negative value of $t_2$. Thus,
we obtain the desired geometry connecting the decelerating universe to the
accelerating universe where its acceleration tends to vanish eventually.

\begin{figure}[t]
  \includegraphics[width=0.6\textwidth]{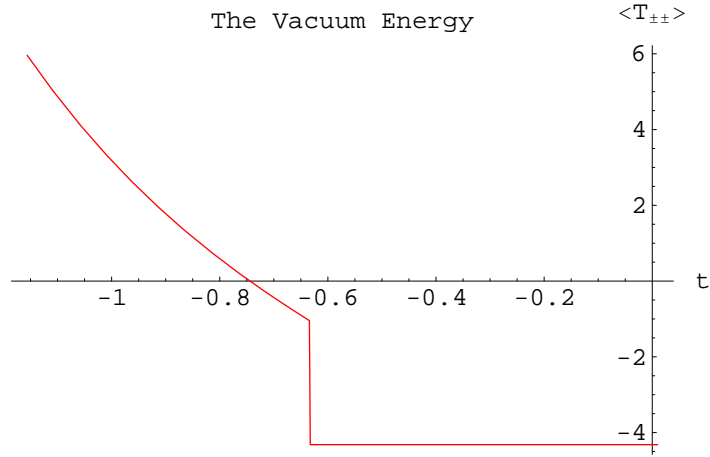}
  \caption{The abrupt drop of the vacuum energy happens at
          $t=t_2$. This figure is plotted for the same constants in 
          Fig.~\ref{fig:R}.}
  \label{fig:T}
\end{figure}

Finally, as for the induced vacuum energy~(\ref{energy:com}), it is
constant, which is explicitly written as
\begin{eqnarray}
  <T_{\pm\pm}> &=& - 2\kappa P_{\Omega_0}^2 \left( 1 +
              \frac{P_{\chi_0}}{P_{\Omega_0}} \right) \nonumber \\
      &=& -2 \kappa P_{\Omega_0}^2 \left( 1 + \tanh (2\kappa\theta_2
              t_2) \right),
         \label{energy:com:t_2}
\end{eqnarray}
for $t > t_2$ by the use of Eq.~(\ref{patch:chi}), which is negative
energy density. On the other hand, the induced vacuum
energy for $t_1 < t < t_2$ is obtained by 
eliminating the constant $\beta$ in Eq.~(\ref{energy:non}) in terms of
Eq.~(\ref{patch:beta}), 
\begin{equation}
  \label{energy:non:t_2}
  <T_{\pm\pm}> = - \kappa P_{\Omega_0} {\rm sech} (2\kappa\theta_2
     t_2) \left[\kappa\theta_2 \sinh (2\kappa\theta_2 t)  
     + 2 P_{\Omega_0} {\rm sech} (2\kappa\theta_2 t_2)
     e^{2\kappa\theta_2 t} \cosh (2\kappa\theta_2 t) \right].
\end{equation}
Note that it is mostly positive where it becomes negative just before $t_2$.
The vacuum energy is jumped down as seen from Eqs.~(\ref{energy:com:t_2}) and
(\ref{energy:non:t_2}), which is essentially due to our assumption of
the noncommutativity using the abrupt step functions.

The noncommutativity represented by modified Poisson brackets 
gives the phase changing from the decelerating to the accelerating 
phase, however, the acceleration does not end and it eventually
diverges. 
Therefore, the noncommutativity should be terminated at a certain time 
after phase changing. So, the finite accelerating region from 
the commutativity is patched up in order to avoid the divergent 
acceleration. Unfortunately, the duration of the noncommutativity 
is expressed by the simplified  step function, which yields the 
jumped down behavior of the energy momentum tensors. Even in 
this simplified assumption for the noncommutative parameter, $\Omega$
and $\chi$ 
in Eqs. (32)-(35) are continuous at the intersection point up to their 
time derivatives. These requirements show that the dilaton field in
Eq. (27) 
is continuous up to their derivative and subsequently the 
metric or scale factor in Eq. (28) is too. On the other hand, 
the continuity of the scalar curvature guarantees the continuity 
of the second derivatives of the metric since it is written 
as $R=8e^{-2\rho}\partial_{+}\partial_{-}\rho$ in two-dimensions. 
Of course, the curvature scalar is not analytic but continuous. 
Our matching condition does not imply the analyticity of the 
curvature scalar but the continuity of them.  
We expect a smooth matching may be possible if we take a smooth 
noncommutativity parameter, which has not been studied in this work. 

The accelerating cosmology naturally appears in the semi-classically
quantized dilaton gravity called the BPP model in the black hole
model~\cite{bpp}. In this cosmological model, the dilaton driven
acceleration is not a weird phenomenon in that the dilaton field plays
an ghost or phantom-like role in terms of its wrong sign kinetic term
in our starting action, which is on the contrary to the conventional
Einstein theory which predicts the deceleration with the ordinary
matter. These two drastically different contents are incorporated in
the present model through the dilaton driven acceleration and the
vacuum energy driven deceleration. The latter in
Eq.~(\ref{energy:non}) is mostly positive, it behaves as an
ordinary matter which contributes deceleration of the universe.

In some sense, the dark energy is originated from the the dilaton in
our model and the description of the decelerating universe becomes
impossible, so that we have considered the
quantum-mechanically induced normal energy which partially compensates
the dark energy in the past. In fact, to obtain the nontrivial
energy-momentum tensor, we have introduced the noncommutative algebra
only for the early time. This technical point is intuitively
compatible with our feeling that the noncommutativity is natural to
apply the early universe instead of the present large universe. 

At first sight, our starting semiclassical action seems to be 
quantized  one more, however, this is not the case since the 
modified Poisson brackets are simply the counterpart of the 
conventional Poisson brackets which are not quantum commutators. 
In the Hamiltonian formulation using the usual Poisson brackets, 
the Hamiltonian equations of motion written in the form of the 
first order with respect to the time can be classically solved, 
then the solutions are exactly same with those of the original 
Euler-Lagrangian equations of motion unless we regard the fields
 as operators. If the fields had been taken as operators by 
decomposing the positive and the negative frequency modes along 
with the normal ordering, then that would be the quantization of 
a quantization. But our modified Poisson brackets just modify the 
conventional (semiclassical) Hamiltonian equations of motion, 
which still result in  the semiclassical solutions, of course, 
they are theta dependent due to the modification of the Poisson
brackets. 
Unfortunately, in our model, we do not know how to obtain theta 
dependent Euler-Lagrangian equations of motion directly from the 
Lagrangian. There may be such a nice Lagrangian formulation 
depending on models case by case as very slowly moving point 
particle in the constant magnetic field or D-branes in a 
constant Neveu-Schwarz two form field studied originally in
Ref. [20]. 
On the other hand, our theta-independent classical and 
semiclassical action do not give the desired phase change of 
acceleration. Thus, the purpose of this modification is 
to find whether the phase changing solution can be obtained or not. 
So, our solution is not the quantized one of the semiclassically 
quantized model but the theta dependent semiclassical solution. 
These theta dependent Poisson brackets can be applied 
at the various level of quantization. Secondly, 
the reason why we applied the modified theta dependent 
Poisson brackets to the semiclassical action (4) instead of 
the original classical action (2) is to use the local 
undetermined function $t_\pm$ in the semiclassical 
version which is related to the vacuum state of the 
quantized matters. It was firstly introduced in Ref. [6] to determine 
the geometry of the black hole, which is absent in the classical
theory.  
The phase change is essentially related to the energy momentum 
tensors, and the fine-tuned classical energy-momentum tensors
 may give the phase changing solution, however, 
it seems to be more or less ad hoc.  However, our 
model is based on the fact that the necessary 
energy and pressure in order for the phase change 
come from the part of quantized matters through $t_\pm$ in the semiclassical theory.    

In our model, there is an initial singularity at $t_1$, which may be
removable in the other quantization scheme. Unfortunately, what is
worse, this model does not contain the transition from the
inflationary era to the decelerating phase in the early universe. So,
it might be interesting to study these problems in this scheme. 

In summary, our model does not describe our
whole genuine universe, though, it seems to be meaningful to suggest
an alternative to show the phase transition from the deceleration
universe to the accelerating universe chronologically through the
analytic model.


\begin{acknowledgments}
 We would like to thank E.~Son for exciting discussions.
This work was supported by the Sogang Research Grant, 20061055.  
\end{acknowledgments}



\begin{thebibliography}{99}

\bibitem{perlmutter} S.~Perlmutter {\it et al.}, Astrophys.\ J.\
  \textbf{517} (1999) 565 [astro-ph/9812133].

\bibitem{spergel} D.~N.~Spergel {\it et al.},  Astrophys.\ J.\ Suppl.\
  \textbf{148} (2003) 175 [astro-ph/0302209].

\bibitem{caldwell} R.~R.~Caldwell, Phys.\ Lett.\ B \textbf{545} (2002)
  23 [astro-ph/9908168].

\bibitem{carroll} S.~M.~Carroll, A.~D.~Felice, V.~D., D.~A.~Easson,
  M.~Trodden, and M.~S.~Turner, Phys.\ Rev.\ D \textbf{71}(2005) 063513
  [astro-ph/0410031].

\bibitem{bbmm} T.~Biswas, R.~Brandenberger, A.~Mazumdar, and
  T.~Multamaki, \textit{Current acceleration from dilaton and stringy
  cold dark matter}, hep-th/0507199.

\bibitem{no} S.~Nojiri and S.~D.~Odintsov, Phys.\ Rev.\ D \textbf{70}
  (2005) 103522 [hep-th/0408170]. 

\bibitem{klm} H.~Kim, H.~W.~Lee, and  Y.~S.~Myung, Phys.\ Lett.\ B
  \textbf{632} (2006) 605 [gr-qc/0509040].

\bibitem{wc} H.~Wei and R.-G.~Cai, Phys.\ Rev.\ D \textbf{73} (2006)
  083002 [astro-ph/0603052].

\bibitem{mrcm} R.~B.~Mann and S.~F.~Ross, Phys.\ Rev.\ D \textbf{47}
  (1993) 3312 [hep-th/9206022];
  K.~C.~K.~Chan and R.~B.~Mann, Class.\ and Quant.\ Grav.\ \textbf{10}
  (1993) 913 [gr-qc/9210015].

\bibitem{gv}  M.~Gasperini and G.~Veneziano, Phys.\ Lett.\ B
  \textbf{387} (1996) 715 [hep-th/9607126].

\bibitem{rey} S.~J.~Rey,  Phys.\ Rev.\ Lett.\ \textbf{77} (1996) 1929
  [hep-th/9605176]. 

\bibitem{bk} S.~K.~Bose and S.~Kar, Phys.\ Rev.\ D \textbf{56} (1997)
  4444 [hep-th/9705061].

\bibitem{ky:dg} W.~Kim and M.~S.~Yoon, Phys.\ Lett.\ B \textbf{423}
  (1998) 231 [hep-th/9706154].

\bibitem{ky:bd} W.~Kim and M.~S.~Yoon, Phys.\ Rev.\ D \textbf{58}
  (1998) 084014 [hep-th/9803081].

\bibitem{cdkz} M.~H.~Christmann, F.~P.~Devecchi, G.~M.~Kremer, and
  C.~M.~Zanetti, Europhys.\ Lett.\ \textbf{67} (2004) 728
  [gr-qc/0407029].
 
\bibitem{cghs} C.~G.~Callan, S.~B.~Giddings, J.~A.~Harvey, and
  A.~Strominger,  Phys.\ Rev.\ D \textbf{45} (1992) 1005
  [hep-th/9111056].
  
\bibitem{rst}   J.~G.~Russo, L.~Susskind, and L.~Thorlacius, Phys.\
  Rev.\ D \textbf{46} (1992) 3444 [hep-th/9206070].

\bibitem{bpp} S.~K.~Bose, L.~Parker, and Y.~Peleg, Phys.\ Rev.\ Lett.\
  \textbf{76} (1996) 861 [gr-qc/9508027].

\bibitem{kv} W.~Kummer and D.~Vassilevich, Phys.\ Rev.\ D \textbf{60}
  (1999) 084021 [hep-th/9811092].

\bibitem{sw} N.~Seiberg and E.~Witten, J.\ High Energy Phys.\
  \textbf{09} (1999) 032 [hep-th/9908142].

\bibitem{bn} G.~D.~Barbosa and N.~Pinto-Neto, Phys.\ Rev.\ D
  \textbf{70} (2004) 103512 [hep-th/0407111]. 

\bibitem{vas} D.~Vassilevich, \textit{Stability of a noncommutative
  Jackiw-Teitelboim gravity}, hep-th/0602095.
 
\bibitem{ko} W.~Kim and J.~J.~Oh, Mod.\ Phys.\ Lett.\ \textbf{15}
  (2000) 1597 [hep-th/9911085].

\bibitem{bc} A.~Bilal and C.~Callan, Nucl.\ Phys.\ B \textbf{394}
  (1993) 73 [hep-th/9205089].

\bibitem{bk:H} A.~Bilal and I.~I.~Kogan, Phys.\ Rev.\ D \textbf{47}
  (1993) 5408 [hep-th/9301119].
\end{thebibliography}
\end{document}